\documentclass[prl,twocolumn,showpacs]{revtex4}
\usepackage{graphicx}
\usepackage[large]{subfigure}
\usepackage{ae}

\newcommand{\rb}{$^{87}$Rb}
\newcommand{\kq}{$^{41}$K}

\newcommand{\kqa}{$^{40}$K}

\newcommand{\ket}[1]{| #1 \rangle}
\renewcommand{\prl}{Phys. Rev. Lett.}
\renewcommand{\pra}{Phys. Rev. A}
\renewcommand{\prb}{Phys. Rev. B}

\begin{document}

\title{Degenerate Bose-Bose mixture in a three-dimensional optical lattice}

\author{J.~Catani, L. De Sarlo, G. Barontini, F. Minardi$^{1,2}$ and
  M. Inguscio$^{1,2}$}

\affiliation{LENS - European Laboratory for Non-Linear
Spectroscopy and Dipartimento di Fisica, Università di Firenze,
via N. Carrara 1, I-50019 Sesto Fiorentino - Firenze, Italy\\
$^1$CNR-INFM, via G. Sansone 1, I-50019 Sesto Fiorentino -
Firenze, Italy\\
$^2$INFN, via G. Sansone
1, I-50019 Sesto Fiorentino - Firenze, Italy}

\begin{abstract}

  We produce a heteronuclear quantum degenerate mixture of two
  bosonic species, \rb\ and \kq, in a three-dimensional optical
  lattice. On raising the lattice barriers, we observe the disapperance
  of the inference pattern of the heavier \rb, shifting toward shallower
  lattice depths in the presence of a minor fraction of \kq. This effect
  is sizable and requires only a marginal overlap between the two
  species.  We compare our results with similar findings reported for
  Fermi-Bose mixtures and discuss the interpretation scenarios
  proposed to date, arguing that the explanation may be linked to the
  increased effective mass due to the interspecies interactions.

\end{abstract}


\date{\today}

\maketitle
Quantum degenerate gases are formidable systems to shine light on
fundamental quantum phenomena occurring at extremely low temperatures,
such as superconductivity and superfluidity. In combination with
optical lattices and scattering resonances, degenerate gases give rise
to strongly correlated systems, enriching even further the breadth of
phenomena that can be directly probed. Indeed, the pioneering
experiment on the superfluid to Mott-insulator transition
\cite{greiner-mi} has shown how physical models long studied in the
field of condensed matter can be realized almost ideally. With two
different atomic species, the wealth of quantum phases grows to a
daunting complexity \cite{pd2s}, only marginally explored by
experiments. Actually, experiments with heteronuclear mixtures in
three-dimensional (3D) optical lattice have been performed very
recently only for Fermi-Bose systems
\cite{krb-esslinger,krb-sengstock}, while Fermi-Fermi and Bose-Bose
mixtures are yet uncharted territory. The importance of mixtures in
optical lattices is hardly overstated: association of dipolar
molecules \cite{dipmol}, mapping of spin arrays \cite{mapspin},
schemes for quantum calculation \cite{qcalc}, and implementation of
disorder \cite{disorder} represent only a few major research lines
that potentially will greatly benefit from such systems. In
particular, Bose-Bose mixtures seem well suited for all these
purposes, provided that collisional losses are adequately suppressed.

\begin{figure}[b]
\begin{center}
\includegraphics[width=\columnwidth]{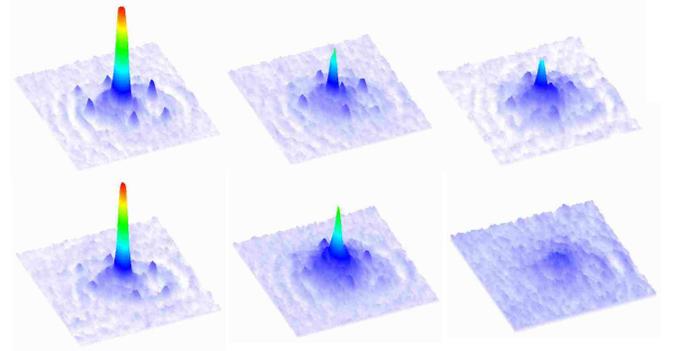}
\caption{(Color online): Interference pattern of \rb\ for $s=6,\,11,\,
  16$ (left to right). Upper row: \rb\ only; lower row: \rb\ mixed
  with \kq\ (not shown). Each panel represents the column density over
  an area $175\,\mu$m $\times\, 135\,\mu$m.}
\label{fig:peaks}
\end{center}
\end{figure}

This work reports the first realization of a degenerate Bose-Bose
heteronuclear mixture in a 3D optical lattice.  Exploiting the large
mass difference, we investigate the regime where one species lies well
in the superfluid domain, while the other exhibits the disappearence
of the interference pattern usually associated with the transition
from a superfluid to a Mott insulator.  We focus only on the
interference pattern and do not probe the shell structure
characteristic of a Mott-insulator for trapped systems
\cite{shell-bloch, shell-ketterle}.  We observe that the presence of
the superfluid enhances the loss of phase coherence of the second
species occurring for increasing lattice strengths. Surprisingly
enough, we find that the effect is sizable (see Fig.~\ref{fig:peaks}),
not only for small fractions of the minority species, but even for
marginal spatial overlap between the two.

%%% Descrizione dell'esperimento 

The experimental setup has been described earlier~\cite{catani-pra,
  luis-pra}. Here we briefly outline the production of a double Bose-Einstein condensate (BEC),
first demonstrated in \cite{doubleBEC-bec2}.  We capture $2\times
10^8$ \rb\ atoms and a variable number of \kq\ atoms in the stretched
$\ket{F=2,m_F=2}$ states in a Ioffe millimetric trap (MMT)
\cite{ruquan-pra,luis-pra} and we start evaporation by lowering the
MMT depth from 5 to 3.5~mK, with final harmonic frequencies
$(\omega_x,\omega_y,\omega_z)=2\pi\times (16.8,202,202)$~Hz
\cite{freq-note}. We then cool \rb\ only by
microwave evaporation, while \kq\ is cooled
sympathetically.  A single sideband, 10~dB
below the carrier, selectively removes \rb\ atoms in the
$\ket{F=2,m_F=1}$ state, which would quickly deplete the \kq\
sample. BEC is achieved first for \kq\ at 250 nK and then for \rb:
condensates contain typically $4\,(2) \times 10^4$ \rb (\kq) atoms, with
Thomas-Fermi radii of 2.4(2.2)~$\mu$m and peak densities of $1.4(0.9)
\times 10^{14}$~cm$^{-3}$. 

In the presence of $2\times 10^3$ \kq\ atoms clearly condensed (inverted
ellipticity after expansion), since no thermal cloud can be detected,
our sensitivity places an upper bound of 50\% on the thermal
fraction of this species: this sets an upper bound on the temperature
of 73~nK. Since the evaporation is stopped at the same level also in the
absence of \kq, the same temperature boundary applies also for \rb\ alone.
At 73~nK, the peak density of the \kq\ thermal atoms is lower than $1/60$
that of the condensate. For this reason, we will
throughout neglect the \kq\ thermal fraction in the following.

The clouds are only partially overlapped: due to differential gravity
sag, the \rb\ condensate lies 3.2~$\mu$m below \kq. Also, numerical
integration of the 3D Gross-Pitaevskii equation (GPE) shows that both
density distributions are deformed and almost completely
phase separated, due to the strong K-Rb repulsion
\cite{topology-riboli}. Indeed, the interspecies scattering length
$a_{\rm K-Rb}=163a_0$ \cite{feshbach-bec2} is much larger than both
$a_{\rm Rb}=99a_0$ \cite{rempe} and $a_{\rm K}=65a_0$
\cite{k41as-ct}, with $a_0$ the Bohr radius.

Once double BEC is achieved, we ramp three retroreflected
optical lattice beams at $\lambda_L=1064$~nm, with frequencies
differing by tens of megahertz,
propagating along the $x,y$, and $z$
axes and focusing at the center of the MMT with waists equal to
(90,180,160)~$\mu$m.  The lattice strength is calibrated by means
of Bragg oscillations and Raman-Nath diffraction \cite{RN-nist},
yielding a systematic uncertainty of 10\%, which must be added to the
statistic uncertainties quoted hereafter.  We choose the duration and
time constant of the exponential turn-on profile, 50 and 20 ms,
respectively, by maximizing the visibility of the \rb\ interference
pattern. We wait for 5 ms with the optical lattice at full power and
abruptly switch off the lattice beams ($< 1\,\mu$s) and the MMT
current ($\sim$ 100$~\mu$s). We image both species after a typical
time of flight of 15 to 20~ms with resonant absorption: the
interference peaks of \rb\ are progressively smeared as the lattice
strength is ramped at higher values.

\begin{figure}[t]
\begin{center}
\includegraphics[width=\columnwidth]{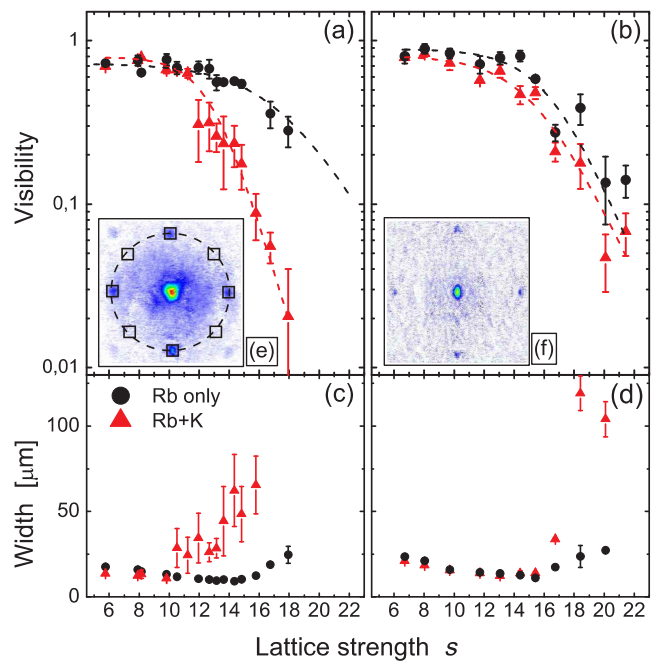}
\caption{(Color online) Visibility and width of the central peak of
  the \rb\ interference pattern, for partial (a,c) and zero (b,d)
  overlap with the \kq\ condensate. In each panel we compare data and
  fit (dashed lines) with \kq\ atoms (red) and without (black),
  i.e. with $N_K=2(1)\times 10^3$ and 0. The insets show: how
  visibility of \rb\ is extracted from images (e) and the interference
  pattern of \kq\ visible for s=20 (f).}
\label{fig:viswidth}
\end{center}
\end{figure}

We measure the width of the central peak and the visibility $v$ of the
interference pattern, defined as \cite{mottvisibility-gerbier}
$v=(N_p-N_d)/(N_p+N_d)$, where $N_p$ and $N_d$ denote, respectively,
the summed weights of the first lateral peaks and of equivalent
regions at the same distance from the central peak along the diagonals
[see Fig.~\ref{fig:viswidth}(e)]. At our lattice wavelength, the
optical potential is the same within 10\% for \rb\ and \kq. However,
due to the their larger mass and scattering length, \rb\ atoms tunnel
less and repel each other more: the combined effect is that \rb\
loses phase coherence at lower lattice power than \kq.

% descrizione dei risultati a 70A

The main experimental result of this work is the observation that the
lattice strength at which the \rb\ interference pattern starts washing
out is greatly shifted not only by a minor admixture of \kq\ atoms,
but also with a marginal spatial overlap. These findings, displayed in
Fig.~\ref{fig:peaks}, augment those of similar experiments carried out
with the mutually attractive \kqa-\rb\ Fermi-Bose mixture
\cite{krb-esslinger, krb-sengstock}.

To quantify the shift of the transition point, we plot the visibility
and the width of the central peak versus the lattice strength in units
of the \rb\ recoil energy $s$ \cite{sk}. We compare data taken with \kq\ and
when \kq\ was not even loaded in the MMT (Fig.~\ref{fig:viswidth}). We
fit the visibility with a phenomenological Fermi function
$v = v_0/[1+\exp (\alpha (s-s_c))]$,
which has the expected flat behavior below the critical lattice
strength $s_c$ and decreases exponentially for $s\gg s_c$. From the fit,
we find that, for $N_{\rm Rb}=3\times 10^4$, the critical $s$ value is
$s_c= 16.8\pm 0.4$ for \rb\ only, while $s_c=12.4\pm 0.3$ with $N_{\rm
  K}=(2\pm 1) \times 10^3$.

Using the formula $\eta=0.696
s^{-0.1} \exp(2.07\sqrt{s}) (a/\lambda_L)$ \cite{mottvisibility-gerbier}, we relate $s$ with the ratio
of interaction to tunneling matrix elements, $\eta=U/(6 J)$, in terms
of the lattice wavelength $\lambda_L$ and the scattering length $a$.
From the fit values, we derive both the critical value $\eta_c$ and
the exponents of the visibility decay for $s\gg s_c$, $v\sim
\eta^{-\nu}$: without \kq\ we find $\eta_c=12.3^{+6.2}_{-4.3}$ and
$\nu=2.3^{+0.6}_{-0.4}$, but with \kq\ we have
$\eta_c=3.9^{+1.6}_{-1.2}$ and $\nu=3.4^{+0.8}_{-0.5}$ (error bars
are dominated by the calibration uncertainty of the lattice strength).

The width of the central peak measures the inverse correlation length:
it signals the transition
with a stark climb at $s \simeq 10$ in the presence of \kq\ and $s \simeq
14$ in the absence thereof [see Figs.~\ref{fig:viswidth}(c) and \ref{fig:viswidth}(d)]. After the
transition, the width increase is steeper in the presence of \kq\ even for
null overlap (Fig.~\ref{fig:viswidth}(d)): this behavior is
unexpected and so far unexplained. While the transition points
detected by the width and the visibility are different, the shift is
the same, $\Delta s_c \simeq 4$.

\begin{figure}[b]
\begin{center}
\includegraphics[width=.8\columnwidth]{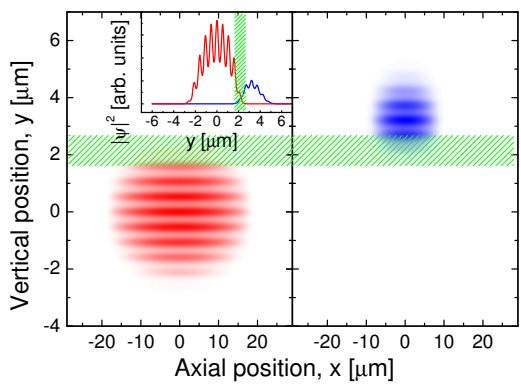}
\caption{(Color online): density distribution of \rb\ (red, left) and
  \kq\ (blue, right) atoms, in the plane $z=0$, calculated by
  numerical integration of the GPE: $N_{\rm Rb}=3\times 10^4$, $N_{\rm
    K}=2\times 10^3$, vertical lattice strength $s=11$. Shaded area
  (green) highlights the overlap, while the inset plot reports density
  profiles along the vertical line through $x=0$.}
\label{fig:gpe}
\end{center}
\end{figure}

Such a shift is surprising, given the little overlap between the two
species.  Numerical integration of the 3D GPE with an $s =11$ vertical
lattice shows that the overlap is restricted to one lattice site out of
11 (see Fig.~\ref{fig:gpe}). Generalizing to a 3D lattice, we expect
that approximately only 10\% of sites are simultaneously filled with
both \rb\ and \kq\ atoms. In order to check that the shift is
genuinely related to the interspecies interaction, we repeat the
experiment with larger differential sag of the two samples. Once the
double-species BEC is achieved, we relax the MMT harmonic frequencies
to $\vec{\omega}=2\pi\times(9.2,108,108)$~Hz, thereby increasing
the vertical separation to 11~$\mu$m, 
so that the overlap is totally negligible even for any
undetected \kq\ thermal cloud ($1/e^2$ radius = 7.7~$\mu$m).
Figures~\ref{fig:viswidth}(b) and \ref{fig:viswidth}(d) show that, although the visibility is
slightly lower in the presence of \kq, the transition point is the same
within our error bars: $s_c=15.8(0.5)$.

\begin{figure}[t]
\begin{center}
\includegraphics[width=.8\columnwidth]{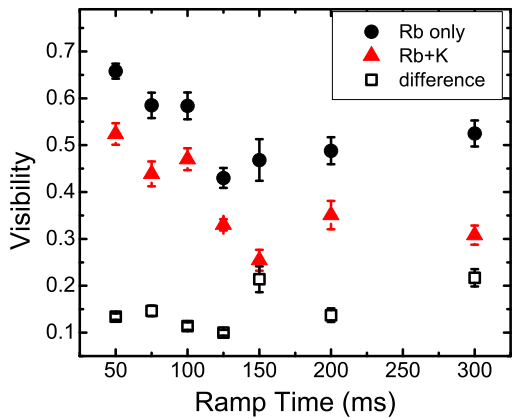}
\caption{(Color online) Visibility of the fringes at $s= 12$ versus
  the duration of the lattice ramp up.}
\label{fig:ramptime}
\end{center}
\end{figure}

As a further test, we investigate the effect of the lattice turn-on
ramps. As mentioned, for \rb\ alone, the visibility is maximum for a
ramp of 50~ms. {\it A priori}, this ramp duration could be insufficient in
the presence of \kq\ impurities. Therefore, we monitor the visibility
versus the ramp duration. The turn-on profile follows essentially an
exponential function with a time constant equal to 0.4 of the total
ramp duration, smoothed at the end. We find that the visibility
decreases for turn-on times longer than 50~ms also in the presence of \kq\
(see Fig.~\ref{fig:ramptime}) and indeed the visibility difference is
roughly constant at 15\%. We conclude that, while in principle one
could expect that in the presence of \kq\ the lattice ramp should be
slower, there is no evidence that this is indeed the case, at least as
far as the loss of visibility is concerned.  Moreover, we do not
observe a significant change in the number of atoms with the duration
of the ramp; hence three-body losses play a negligible role,
consistently with the small overlap of the two clouds.
 
As mentioned, our experimental findings extend those of
Refs.~\cite{krb-esslinger, krb-sengstock} to \textit{bosonic}
impurities and help to shine some light on the induced loss of
coherence, whose origin has so far remained unclear, if not
controversial. G\"{u}nter \textit{et al.} report that the
transition point, extracted from the visibility data with a fitting
procedure equivalent to ours and expressed in terms of $\eta_c$,
shifts from 6.5 to 2.5 for an 8\% admixture. Both values are lower
than ours. Ospelkaus \textit{et al.} instead quantify the shift in
terms of $s_{c}$, as we do, and fit the visibility with the Fermi
function. The reported shift $\Delta s_{c}=3(1.5)$ for a 7~\% fraction
of impurities agrees with our result. With the same data, but measured
from the width of the central peak, the transition point shifts from
$\eta_c=9$ to 4.5 \cite{krb-esslinger} and by $\Delta s_{c}=1(1)$
\cite{krb-sengstock}.

Several different scenarios have been proposed to explain the
impurities-induced loss of coherence. We discuss here the impact of our
experiment on these scenarios.

First, we notice that one would naively expect that strongly
interacting impurities would locally increase the \rb\ atom filling
factor and hence the chemical potential, by expelling or attracting
them. However, increasing the chemical potential leads to an
enhancement of the superfluid fraction, which is indeed the result
predicted by Ref.~\cite{dipmol}, but opposite to what we
observe. Ospelkaus \textit{et al.} argue that fermionic atoms act
like randomly localized scatterers for the boson order parameter,
thereby splitting it into isolated superfluid domains unable to maintain
a single coherent phase throughout the sample
\cite{krb-sengstock}. The suggested phenomenon, reminiscent of
Anderson localization, depends crucially on the fermionic nature of
the \kqa\ impurities, leading to localization. However, bosonic \kq\
atoms are expected to localize less, not more, than \rb.
%, due to their lighter mass
Therefore, the interpretation based on
Andersonlike localization does not hold in our case and might
be unnecessary for the Fermi-Bose experiments.

The loss of phase coherence could follow from heating of the two
clouds when the lattice ramps up. In addition to technical noise, which plays
a role only at longer time scales, we consider the effect of the
thermodynamic transformation involved.  With two species, the
effective masses change at different rates and thermal equilibrium
requires an interspecies redistribution of entropy. In our experiment,
this increases the entropy of \rb\ and therefore its thermal
component.  Quantitatively, due to the small overlap, we neglect the
interspecies interaction energy and calculate the total entropy as the
sum of two independent contributions.  At $s = 0$, we calculate the
contribution of each component from the formula for the total energy:
$E/N= 3 k_B T_c \zeta (4) t^4/\zeta (3) + \mu (1-t^3)^{2/5} (5+16
t^3)/7$, with $t=T/T_c$, where $T_c$ and $\mu$ denote the critical
temperature and the chemical potential \cite{review-tn}.  At $s=20$
the contribution of \kq\ is given by the above formula (replacing the
atomic mass with the effective mass of the lowest energy band), while
for the contribution of \rb\ we take 
that of a deep Mott insulator
\cite{ho}: $S_{\rm shell}/k_B= 32 \pi^3
k_B T \sum_j R_j /(3 m \omega^2 \lambda_L^3)$, where $R_j$ and $m$
indicate the radii of the superfluid layers of the Mott shell
structure and the atomic mass.  For an initial temperature of 73~nK
(\rb\ thermal fraction equal to 0.21), the presence of \kq\ increases
the \rb\ entropy and hence its thermal fraction, by 20\%.
As a consequence, the \rb\ condensate
fraction decreases from 79\% to 75\% and the visibility by
approximately the same amount \cite{gerbier-cm}. We conclude that this
is a minor effect.

A plausible interpretation of our findings invokes the dressing of
\rb\ atoms by \kq: naively, dressed \rb\ atoms weigh more, therefore
tunnel less. This argument is made quantitative by Bruderer {\it et
  al.} \cite{jaksch-polaron}, who analyze the generation of polarons
in a Bose-Bose mixture, i.e., quasiparticles composed of localized
atoms (\rb) plus a cloud of superfluid phonons of the other species
(\kq). The polarons have two different effects: at zero temperature they
reduce the tunneling rate of \rb\ atoms according to the picture
outlined above; at finite temperature, \kq\ phonons undergo incoherent
scattering, further hampering the hopping of \rb. According to
Ref.~\cite{jaksch-polaron}, with the parameters of our experiment at
$s =10$, the energy scale relevant for the onset of the incoherent
scattering is $E_p/k_B \approx 1$ nK and therefore it is comparable
to and lower than the estimated lattice temperature, ranging from 10 to 30
nK, while the tunneling suppression at $T=0$ appears negligible.  The
analysis carried out in Ref.~\cite{jaksch-polaron} relies on suitably
weak interspecies interactions. In our experiment the effective
interspecies interactions are reduced by the low density of \kq\ in
the overlap region but the approximations of
Ref.~\cite{jaksch-polaron} are only weakly satisfied.  The order of
magnitude of the above value, however, proves that the inhibition of
tunneling due to incoherent scattering of phonons deserves a deeper
analysis.

Finally, we address the question of why the shift is substantial even
with the small overlap shown by Fig.~\ref{fig:gpe}. We conjecture
that, owing to the inhomogeneous density, the loss of coherence starts
from the borders of the \rb\ condensate where the filling factor is lower;
the overlap with \kq, although small, occurs in this critical
region. 
Alternatively, one could call on collisions occurring during the
expansion: since we do not observe any degradation of the \rb\
Raman-Nath diffraction pattern caused by the presence of \kq,
time-of-flight collisions between the two species seem unable to
significantly alter the \rb\ momentum distribution, at least in the
superfluid regime. 

In summary, we have created  a degenerate Bose-Bose
mixture in a 3D optical lattice and studied the influence of a minor
admixture of superfluid \kq\ on the loss of \rb\ phase coherence.
We find that this loss is furthered in
the presence of \kq, even if the two clouds overlap only at their
boundaries.  Our results agree qualitatively with similar observations
carried out with a Fermi-Bose mixture, notwithstanding the different
statistics of impurities and the opposite sign of the strong
interspecies interactions.  Following a recent theoretical analysis,
we find that the activation energy of polarons is of the same order
as the estimated temperature in our optical lattice.  Our degenerate
Bose-Bose mixture will also be studied in the double Mott-insulator
regime, promising for the formation of stable molecules. To this end,
Feshbach resonances predicted at moderate magnetic fields are
instrumental to control the interspecies interactions.

This work was supported by Ente CdR in Firenze, INFN through the project
SQUAT-Super, and EU under Contracts No. HPRICT1999-00111 and
No. MEIF-CT-2004-009939. We thank P. Maioli, who started this experiment
with us, and all the members of the Quantum Degenerate Gas group at
LENS for fruitful discussions.


\begin{thebibliography}{99}
\bibitem{freq-note} Hereafter we quote frequencies referred to \rb;
  \kq\ frequencies are higher by a factor
  of 1.46.%$\sqrt{m_{\rm Rb}/m_{\rm K}}\simeq 1.46$
\bibitem{sk} At a given lattice power, the $s$ parameter is a factor
  of 2.4 lower for \kq.
\end{thebibliography}
\end{document}